 \documentclass[sigconf,table]{acmart}

\AtBeginDocument{%
  \providecommand\BibTeX{{%
    \normalfont B\kern-0.5em{\scshape i\kern-0.25em b}\kern-0.8em\TeX}}}

\copyrightyear{2021}
\acmYear{2021}
\setcopyright{acmcopyright}
\acmConference[CHI '21]{CHI Conference on Human
Factors in Computing Systems}{May 8--13, 2021}{Yokohama, Japan}
\acmBooktitle{CHI Conference on Human Factors in Computing Systems (CHI '21),
May 8--13, 2021, Yokohama, Japan}
\acmPrice{15.00}
\acmDOI{10.1145/3411764.3445801}
\acmISBN{978-1-4503-8096-6/21/05}

\begin{document}

\title{Effect of Gameplay Uncertainty, Display Type, and Age on Virtual Reality Exergames}


\author{Wenge Xu}
\affiliation{%
  \institution{Xi'an Jiaotong-Liverpool University}
  \city{Suzhou}
    \state{Jiangsu}
  \country{China}}
\email{wenge.xu@xjtlu.edu.cn}

\author{Hai-Ning Liang}
\authornote{Corresponding author.}
\affiliation{%
  \institution{Xi'an Jiaotong-Liverpool University}
  \city{Suzhou}
  \state{Jiangsu}
  \country{China}}
\email{haining.liang@xjtlu.edu.cn}

\author{Kangyou Yu}
\affiliation{%
  \institution{Xi'an Jiaotong-Liverpool University}
  \city{Suzhou}
    \state{Jiangsu}
  \country{China}}
\email{kangyou.yu18@student.xjtlu.edu.cn}

\author{Nilufar Baghaei}
\affiliation{%
  \institution{Massey University}
  \city{Auckland}
  \country{New Zealand}}
\email{n.baghaei@massey.ac.nz}

\renewcommand{\shortauthors}{Xu et al.}

\begin{abstract}
 Uncertainty is widely acknowledged as an engaging gameplay element but rarely used in exergames. In this research, we explore the role of uncertainty in exergames and introduce three uncertain elements (false-attacks, misses, and critical hits) to an exergame. We conducted a study under two conditions (uncertain and certain), with two display types (virtual reality and large display) and across young and middle-aged adults to measure their effect on game performance, experience, and exertion. Results show that (1) our designed uncertain elements are instrumental in increasing exertion levels; (2) when playing a motion-based first-person perspective exergame, virtual reality can improve performance, while maintaining the same motion sickness level as a large display; and (3) exergames for middle-aged adults should be designed with age-related declines in mind, similar to designing for elderly adults. We also framed two design guidelines for exergames that have similar features to the game used in this research.
\end{abstract}




\maketitle

\section{Introduction}
Motion-based exergames, a combination of “motion-based exercise” and “gaming”, is a promising approach to encourage regular exercise, especially for unmotivated or inactive target groups \cite{bogost_rhetoric_2005,sinclair_considerations_2007}. Previous literature has shown the benefits of playing motion-based exergame, which include but are not limited to enhanced postural stability \cite{sheehan_effects_2013}, muscle strength \cite{soares_serious_2016}, and working memory \cite{eggenberger_does_2015}. Because of the potential of these exergames in eliciting health benefits, much work has been conducted with different age groups (including children \cite{hernandez_designing_2013}, young individuals \cite{xu_studying_2020}, and older adults \cite{gerling_full-body_2012}). 

Age-related declines are common in older adults (i.e., aged 65 and above) and middle-aged adults (i.e., aged 45 to 65) as previous studies show that reductions (e.g., cognitive abilities) could start even before the age of 50 \cite{ferreira_cognitive_2015,verhaeghen_meta-analyses_1997}. These age-related declines affect the elderly’ game performance and experience and could also affect middle-aged adults in a similar way. Although there have been some attempts to understand whether middle-aged adults could obtain the same health benefits from playing videogame as elderly adults \cite{rosney_exergaming_2018,xu_health}, there is very limited research on exploring the performance and experience of middle-aged adults.

Designing an enjoyable and effective exergame is challenging. Studies \cite{berkovsky_physical_2010,ioannou_virtual_2019,barathi_interactive_2018} have been conducted to improve the motivation and experience of these games. For instance, Ioannou et al. \cite{ioannou_virtual_2019} proposed a virtual performance augmentation method for exergames and found that it increased players’ immersion and motivation. Barathi et al. \cite{barathi_interactive_2018} implemented an interactive feedforward method to an exergame and found that it improved players’ performance. 

One factor that has been widely applied in games is uncertainty, which has long been recognized as a key ingredient of engaging gameplay \cite{costikyan_uncertainty_2013, power_lost_2019,caillois_man_2001,johnson_unpredictability_2018}. Costikyan \cite{costikyan_uncertainty_2013} argues that games require uncertainty to hold players’ interest and that the struggle to master uncertainty is central to games’ appeal. Most importantly, he suggested that game designers can harness uncertainty to frame gameplay's design. Several game designers and researchers have tried to identify uncertainty sources that can lead to a good gameplay experience \cite{Thomas_uncertainty,juul_half-real_2011,tekinbas_rules_2003,dekoven_well-played_2002}. Drawing on many of these sources and practical experience, Costikyan \cite{costikyan_uncertainty_2013} listed an influential categorization of eleven sources of uncertainty found or can be used in games. Recently, Kumari et al. \cite{kumari_uncertainty} presented a grounded theory of uncertainty sources which can partially map onto existing taxonomies, especially Costikyan's \cite{costikyan_uncertainty_2013}, providing converging evidence of the validity of Costikyan's categorization of uncertainty sources. Although uncertainty is recognized as a core component of the gaming experience, there is relatively little research that has looked specifically into the effect of uncertainty in games, especially exergames. Based on uncertainty sources identified in \cite{costikyan_uncertainty_2013}, in this research, we propose the use of three uncertain elements for exergames, that cover four sources of uncertainty, and evaluate their effect in exergames on performance, game experience (and sickness when implemented in virtual reality), and exertion levels.

Given the recent emergence of affordable virtual reality (VR) head-mounted displays (HMDs), VR exergames have been gaining rapid attention \cite{barathi_interactive_2018,ioannou_virtual_2019,virusboxing}. For instance, VR exergames are useful in promoting physical activity in sedentary and obese children \cite{rizzo_virtual_2011}, especially to increase their motivation to exercise \cite{plante_does_2003,mestre_does_2011}. Existing literature has outlined that there are additional benefits of playing motion-based exergames in VR than non-VR. In VR, players could achieve a higher exertion and experience a game more positively in areas like the challenge, flow, immersion and a lower negative affect \cite{xu_studying_2020}. However, a major drawback is that VR might lead to a higher level of simulator sickness, which must be taken into account during the design process to mitigate its effects.

The aim of our research is to explore the effect of uncertain versus certain elements and VR versus a typical TV large display (LD) on two main player groups of exergames regarding their game performance, experience, and exertion. In this paper, we first introduce \textit{GestureFit}, the game we developed for this research. We describe the rules and logic behind it, the game procedure, and risk control for middle-aged adults. We then present the study we conducted to investigate the effect of display type and game condition, focusing on differences between young adults and middle-aged adults. We then report the results and present a discussion of our findings that are framed based on existing literature. Two main design guidelines derived from the results are then proposed, followed by the conclusions.

The contributions of the paper include: (1) an empirical evaluation of the effects of display type and game condition on exergame performance, experience, and exertion between young and middle-aged adults; (2) a set of uncertain elements that can help increase the exertion level for motion-based exergames; and (3) two recommendations that can help frame the design of motion-based exergames to contain uncertain gameplay elements and how to motivate middle-age and older adults to engage with exergames more meaningfully.

\section{Related Work}
\subsection{VR and Non-VR Motion-based Exergames}
Many motion-based exergames have been developed for non-VR displays since the introduction of Kinect. A typical motion-based exergame requires players to move their body or perform certain gestures to interact with the game world. For instance, in \textit{GrabApple} \cite{gao_acute_2012}, users need to jump or duck to pick up apples; they also need to move around to locate them but also avoid touching other objects, like bombs. In a game reported in Gerling et al. \cite{gerling_full-body_2012}, users need to perform static and dynamic gestures to grow plants and flowers and catch birds. In \textit{Sternataler} \cite{smeddinck_adaptive_2013}, players use their hands to collect stars that appear sequentially in some predefined paths. 

Recent advances and the growing popularity of VR HMDs have created a substantial demand for motion-based exergames. For instance, games like \textit{Virtual Sports}\footnote{https://www.vrgamerankings.com/virtual-sports} for the HTC VIVE allow a user to play sports with his/her full body in fully immersive virtual environments. In another commercial game, \textit{FitXR}\footnote{https://fitxr.com/}, the users need to jab, weave, and do uppercuts following rhythmic music. In the research exergame \textit{KIMove} \cite{xu_assessing_2019}, the players need to move their hands to hit fruits floating in midair and use their feet to step on cubes moving towards them on the ground. In \textit{GestureStar} \cite{xu_studying_2020}, users need to perform 6 different gestures to eliminate the objects, like cubes, flying towards them. 

Previous research has reported inconsistent findings when looking at the effect of display type on gameplay experience and performance. Xu et al. \cite{xu_studying_2020} suggested that players achieved a higher exertion and experienced a game more positively in VR than LD. However, they also found that VR could lead to a higher level of simulator sickness. Results from \cite{xu_assessing_2019} suggested that there was no effect of display type on gameplay performance and experience. Therefore, we have included this factor in our experiment to investigate it further and provide more insights.

\subsection{User Experience in Exergames}
Exergames integrate physical activity to engage players \cite{mueller_designing_2011}. Because findings from other types of games may not be applicable to exergames \cite{monteiro_evaluating_2018,xu_studying_2020}, efforts have been focused on studying user experience in exergames. For instance, it is reported in \cite{xu_assessing_2019} that task mode (single- and multi-tasking) could affect users’ exergame experience; in particular, multi-tasking could not only make the game more challenging and cause a higher sickness, but also lead to worse performance than single-tasking. Koulouris et al. \cite{koulouris_me_2020} investigated the effect of customization and identification in a VR exergame, and found that customization significantly increased identification, intrinsic motivation, and performance in the exergame. Further, playing pose (i.e., standing and seated), performance augmentation (i.e., enabling players with superhuman capabilities in the virtual game) could also affect the gameplay experience (e.g., sickness) \cite{ioannou_virtual_2019,xu_results_2020}. On the other hand, although uncertainty is a crucial element in gameplay, it is underexplored in exergames. It is this reason that we are interested in studying the effect of uncertainty in exergames for both immersive VR and large displays.

\subsection{Design Elements of Exergames}
Several design guidelines have been proposed by researchers in HCI and sport sciences for designing more attractive and effective full-body motion-based exergames \cite{MARSHALL20161,sequra,hardy_framework_2015}. According to these, to design a playful exergame experience, designers should focus on (1) the player’s body (movement concept), (2) the mediating controller technology (transferring movement input into the virtual world and providing feedback), and (3) the game scenario (audio-visual and narrative design and feedback)  \cite{Anna_Exergame_Related_Work}.

\subsubsection{The Player's Body}
After criticizing existing exertion games and commercial exergames, Marshall et al. \cite{MARSHALL20161} proposed three design strategies based on the idea of movement, which are (1) the design of exertion trajectories (e.g., to create a trajectory across individual play sessions for skill-learning that takes into account players' cognitive load and the exertion patterns), (2) design for, with, and around pain (e.g., celebrating positive pain), and (3) design leveraging the social nature of exertion (e.g., players to be surrounded by other players like friends and family members or game enthusiasts).

\subsubsection{The Mediating Controller Technology}
Studies have suggested that the participation of the body is a crucial variable not only in the efficacy of exergames in affecting users' emotional experience \cite{VARA2016267}, but also in improving user experience, energy expenditure, and intention to repeat the experience \cite{KIM2014376}. To achieve these positive gaming experiences, body-centered controllers should be designed to serve as an additional physical playground, so that they can be easily integrated into players’ body scheme \cite{PASCH200949} and provide a balance of guided and free movements \cite{Anna_Exergame_Related_Work}.

\subsubsection{The Game Scenario}
Exergame should involve specific preferences for game mechanics, levels, visuals, audio, and narrative. This requirement will unavoidably make it essential to involve the target group in the design process from the start \cite{Anna_The_exercube,AnnaDesigning2018}. The literature offers suggestions for key elements of game scenarios. For instance, games should include an immediate celebration of movement articulation by providing direct and constrained amounts of feedback \cite{Mueller_movement_2014}. Also, games should involve achievable short-term challenges to foster long-term motivation and help players identify rhythm in their movements, for example, by setting movements that are mapped to specific sounds and visualizing previous and upcoming movements \cite{Mueller_movement_2014,Muller_Exertion_2016}. It is also important to provide a challenge that matches individual skill levels, for instance, balancing the challenge level by monitoring the player's heart rate \cite{muller_balancing}.

\subsection{Uncertainty in Games}
Caillois \cite{caillois_man_2001} says that the outcome of a game should be uncertain for it to be enjoyable. Similarly, Costikyan \cite{costikyan_uncertainty_2013} argues about the importance of uncertainty in the overall game experience and has developed an influential categorization of 11 sources of uncertainty within games. Typical uncertainty sources are (1) Performative uncertainty: uncertainty of physical performance (e.g., hand-eye coordination); (2) Solver’s uncertainty: weighting a group of options against potential outcomes; (3) Player unpredictability: not knowing how the opponents/teammates will act; (4) Randomness: uncertainty emanating from random game elements. Recently, Kumari et al. \cite{kumari_uncertainty} developed an empirically-based grounded taxonomy of seven sources of uncertainty across the input-output loop that involves the game, the player, and their interaction in an outcome. This taxonomy partially maps onto existing taxonomies, especially the one proposed by Costikyan \cite{costikyan_uncertainty_2013}. This, in turn, provides further evidence of its validity. Hence, in this research, we used Costikyan's sources of uncertainty to guide the design of the uncertainty elements in our exergame. 

To explore the effects of uncertainty in exergames, we applied three uncertain elements in an exergame we developed: (1) \textit{False}-\textit{Attacks}: this concept is originally from sports (e.g., basketball) and has been applied widely in sports videogames (e.g., NBA 2K series). (2) \textit{Misses}: this concept has been widely used in games (e.g., Dungeon \& Fighter) where an attack hits the opponent but is counted as a miss by the system. (3) \textit{Critical Hits}: this concept has also been widely used in games (e.g., Dungeon \& Fighter). When a critical hit happens, the player issuing the hit causes more damages to the opponent that a normal successful blow.

\subsection{Game Experience for Different Age Groups }
Users from different age groups often perceive gameplay elements differently\textemdash for instance, what is motivating for one group may not be so for another. Motivations can change with age: fantasy is a powerful motivational factor in younger children \cite{greenberg_economic_1999}, whereas competition and challenge-related motives are stronger in older children and adolescents \cite{sherry_video_2006}. Young adults are more motivated by rewarding experiences, while older adults are more inspired by perceived benefits to their health \cite{subramanian_assessing_2019}. Young adults tend to prefer visually appealing graphics and music that fit the theme and nature of the game, but older adults pay more attention to the feedback that helps them complete a game \cite{subramanian_assessing_2019}. Furthermore, there is an increased appreciation for the enjoyment that a game brings, greater satisfaction for autonomy, and decreased competence as users age, especially after a certain threshold \cite{birk_age-based_2017}. In other words, young adults prefer exergames that allow them to challenge themselves physically and cognitively, but older adults preferred exergames that are fun to play and are beneficial to their health \cite{subramanian_assessing_2019}. 

Gajadhar et al. \cite{gajadhar_shared_2008} investigated the social elements of gameplay for young adults. They found that gameplay is most enjoyable when gamers are co-located, less satisfying in mediated co-play, and the least enjoyable in virtual co-play. However, these three social contexts (virtual, mediated, and co-located co-play) do not positively influence older users like younger adults \cite{blocker_gaming_2014,gajadhar_out_2010}. Gerling et al. \cite{gerling_is_2013} explored the effect of sedentary and motion-based control tasks in games (such as pointing and tracking) for older adults and younger adults, and found that older adults performed worse than young adults.

There is a large body of work on the experience of children \cite{andries_designing_2019,duh_narrative-driven_2010,eriksson_using_2019} and young adults \cite{xu_results_2020,xu_assessing_2019,xu_studying_2020}, and older adults \cite{de_schutter_never_2011,de_schutter_designing_2010,gerling_full-body_2012} with videogames. However, there is only limited attention given to middle-aged players. Previous research suggested age-related declines could start when people are in their mid-age; for instance, age-related memory impairment and executive dysfunction can be found in people before they reach 50 \cite{ferreira_cognitive_2015,verhaeghen_meta-analyses_1997}. Middle-aged adults suffer from several age-related declines, including but not limited to lower working memory \cite{meguro_nature_2000}, grip strength \cite{kozakai_sex-differences_2016}, and muscle mass \cite{brown_complexity_1996}. Given this above research, our work involves two groups, young adults (18-30) and middle-aged adults (45-65), to explore the effect of age on exergames.

\begin{table*}
  \caption{Features and requirement for each move by the player\textsuperscript{a} and the monster\textsuperscript{b}.}
  \label{tab:freq}
  \begin{tabular}{p{0.1\textwidth} p{0.86\textwidth}}
    \toprule
    Name & Description of the move\\
    \midrule
    \textit{Kick}\textsuperscript{a} & An attack move that inflicts 10 hp damage to the opponent in the kicking direction and requires a 3-second cooldown.\\
    \textit{Punch}\textsuperscript{a,b} & An attack move that inflicts 10 hp damage to the opponent on the punching direction and requires a 3-second cooldown. \\
    \textit{Zoom}+\textit{Kick}\textsuperscript{a} & A ranged attack move that inflicts 30 hp damage to the opponent in that attack range (1m) and requires a 5-second cooldown.\\
    \textit{Squat}\textsuperscript{b} & A ranged attack move that deals 30 hp damage and requires a 5-seconds to cooldown.\\
    \textit{Zoom}+\textit{Squat}\textsuperscript{a}& A defense move that releases a sphere to protect the user for 2 seconds and heals 20 hp if it could successfully defend the player from the monster’s attack. This move requires a 3-second cooldown.\\
  \bottomrule
\end{tabular}
\end{table*}

\section{GestureFit: A Gestured-based Game}
The game was implemented in Unity3D with the Oculus Integration plugin\footnote{https://assetstore.unity.com/packages/tools/integration/oculus-integration-82022} and the Kinect v2 Unity plugin\footnote{https://assetstore.unity.com/packages/3d/characters/kinect-v2-examples-with-ms-sdk-and-nuitrack-sdk-18708}.

\subsection{Rules and Logic}
The design of our game was inspired by \textit{Nintendo Ring Fit Adventure}\footnote{https://www.nintendo.com/games/detail/ring-fit-adventure-switch/}. The goal of the game is for the player to stay alive and defeat a monster three times. To do this, the player needs to perform gestures to make attacks against the monster and defend themselves from being attacked by it. The player begins with 100 health points (HP) while the monster has 500 HP. The monster or player dies when their HP reaches 0. Both the monster and the player have 3 lives. The monster could move leftward or rightward within a 2-meter range prior to its game starting position. Players’ lateral movement is limited so that they are always within the operational tracking range \cite{ioannou_virtual_2019,xu_studying_2020}. The game is designed to take this into account so that the gameplay experience is not affected. Both visual and audio feedback is provided to give a fuller range of sensory experience to players.

\subsubsection{Selected Gestures and Corresponding Attack/Defense Moves}
There are three attack moves and one defense move. All moves can be released by performing their corresponding gestures. These four moves are (i) \textit{Kick}: kicking using any leg, (ii) \textit{Punch}: single hand punching, (iii) \textit{Zoom}+\textit{Kick}: kicking using any leg and leaning arms forward and stretching them out, and (iv) \textit{Zoom}+\textit{Squat}: performing a squat and leaning arms forward and stretching them out. The selected gestures were chosen based on design recommendations from previous studies on young adults \cite{xu_studying_2020} and older adults \cite{gerling_full-body_2012}. Table 1 lists pre-defined features and their requirements.

\subsubsection{The Use of Uncertainty}
The uncertain condition includes three uncertain elements, which covers four uncertainty sources \cite{costikyan_uncertainty_2013}: 



\begin{itemize}
\item{\verb|False-Attacks|}: There is a 20\% chance that the monster would perform a false-attack (which lasts around 0.8 seconds) when the system triggers an attack-related animation to trick the player into performing the defense move. False-attacks cover the following uncertainty sources: a) \textbf{Performative uncertainty}: our game challenges eye-body coordination (i.e., would the players be able to cancel their defense move when they realize the monster is performing a false-attack?), b) \textbf{Solver’s uncertainty}: it is concerned with whether performing or not performing a defense move against potential outcomes (i.e., wasting a defense move to a false-attack or being successful in defending from an actual attack), and c) \textbf{Player unpredictability}: this is about the uncertainty of the opponent’s movements (e.g., whether it is a false or real attack). 
\item{\verb|Misses|}: There is a 10\% chance that the player’s or monster’s attack would be regarded as a miss even if it hits the opponent. \textbf{Randomness}: misses act as a random element in the game. 
\item{\verb|Critical Hits|}: There is a 10\% chance that the player’s or monster’s attack could be a critical hit, which would deal 50\% more damage than a normal attack move. \textbf{Randomness}: critical hits act as another random element in the game.
\end{itemize}

The only difference between the certain and non-certain conditions is that the former does not include the above three uncertain features.

\subsubsection{Monster Attack Design}
In both conditions, the monster would perform an action every 2 sec. In the certain condition, if any attack skill is available, there is 80\% chance that the action is an attack (either 100\% for the only skill that is available or 50\% for each skill that is available); otherwise, it is a walk. The uncertain condition also follows this attack mechanism; the only difference is that if an attack skill is available, there is 80\% chance the action is attack-related (i.e., 8/10 = a real attack, 2/10 = a false attack).

\subsection{Game Procedure}
The game starts with a training (warm-up) phase (see Figure 1a-b), where the player needs to use attack and defense moves. The order of the moves required for the player to perform is \textit{Kick}, \textit{Punch}, \textit{Zoom}+\textit{Kick}, \textit{Zoom}+\textit{Squat}. For attack moves, the player needs to perform the corresponding gesture, and its attack must damage the monster twice before proceeding to the next move. For the defense moves, the player must successfully defend themselves from the monster's attacks twice to finish the training. The player needs to perform a \textit{Zoom} gesture between each move training to switch to the next move training. After the training phase, the player needs to perform another \textit{Zoom} gesture to start the gameplay phase. 

\begin{figure*}
  \centering
  \includegraphics[width=\textwidth]{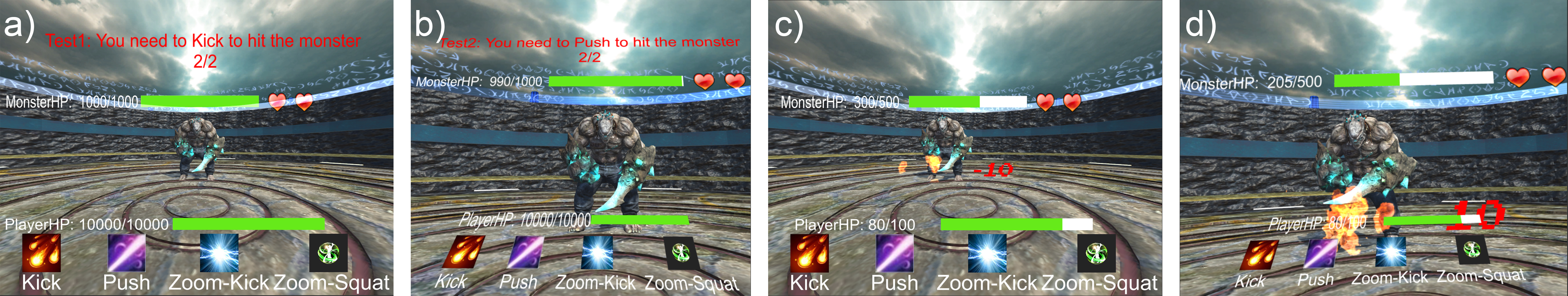}
  \caption{Screenshots of \textit{GestureFit}: (a) LD training phase, (b) VR training phase, (c) LD gameplay phase, and (d) VR gameplay phase. All variables are the same in all versions except in VR the player information is slightly tilted.}
  \Description{Game procedure LD and VR versions}
\end{figure*}

During the gameplay phase (see Figure 1c-d), players need to perform the gestures to attack and defend themselves. If the players have no HPs, they need to perform \textit{Zoom}+\textit{Squat} five times to regain life and perform \textit{Zoom} once to confirm they are ready to return. If the monster has no HPs, the game will play an animation of the monster falling to the ground and is destroyed. After a 5-second wait, the monster uses its second or third life and the game re-starts. The game ends when the monster or the player has no lives and HPs left.

\subsection{Risk Control for Middle-aged Adults}
We controlled the risk, if any, to a minimal level. As pointed in \cite{AnnaDesigning2018,Anna_The_exercube}, having users involved in the development process is useful. As such, for our game prototype, we had two middle-aged adults frequently involved during the development process to test the gestures’ suitability, tune parameters (e.g., cooldown time, shield protection’s duration) and ensure accurate and meaningful execution of movements. The selected gesture worked quite well since all middle-aged participants had no issues performing them during the experimental gaming sessions (as our results would show; more on this later).

Besides, we minimized any risks by (1) making a first-person viewing perspective game so that players can see their motions, (2) limiting the number of monster’s attack skills and having gaps in its attacks, (3) restricting players’ position, (4) allowing them 5 sec rests after they took a monster’s life, (5) allowing them to rest as much as they want after they lost one life, and (6) displaying information (user's skills, player's HP, and monster's HP) in front of the users without the need for additional head movement.

\section{Experiment}
\subsection{Experiment Design and Outcome Measures}
The experiment followed a 2 × 2 within-subjects design with two within-subjects factors: Display Type (DT: VR and LD) and (2) Game Condition (GC: certain and uncertain). The order of DT × GC was counterbalanced in the experiment.

To determine participants’ task performance, we collected the following (1) completion time on each of the three lives of the monster; (2) success rate of each move; and (3) the total number of each type of gestures performed.

Participants’ experience was measured with Game Experience Questionnaire (GEQ) \cite{ijsselsteijn_game_2008} and Simulator Sickness Questionnaire (SSQ) \cite{kennedy_simulator_1993}. We used the 33-item core module of the GEQ to measure game experience, which consists of seven components: competence, immersion, flow, tension, challenge, negative affect, and positive affect. Simulator sickness was assessed using the 16-item SSQ, which produces 3 measures of cybersickness (nausea, oculomotor, and disorientation).

Exertion was evaluated by (1) the average heart rate (avgHR\%) expressed as a percentage of a participant’s estimated maximum heart rate (211-0.64$\times$age) \cite{newHR}, (2) calories burned, and (3) Borg RPE 6-20 scale \cite{borg_psychophysical_1982}.

We measured the acceptability of the uncertain elements used in our games with three questions: “\textit{I like the design of the false-attacks}”, “\textit{I like the design of attacks that could be missed by chance}”, and “\textit{I like the design of attacks that could be a critical hit by chance}”. The questions followed a 1-7 Likert scale, with 1 indicating “extremely disagree” and 7 indicating “extremely agree”.

After completing the above questionnaires, we conducted a semi-structured interview for participants with the following open-ended questions: “\textit{Overall, what did you think about the game}?”, “\textit{What did you like about the game}?”, “\textit{What did you not like about the game}?”, “\textit{Was there anything more difficult than you expected in the game}?”, and “\textit{Was there anything more confusing than you expected in the game}?” \cite{drachen_games_2018}. Answers were recorded and transcribed in text and later analyzed by two of the researchers following an informal, simplified inductive open coding approach \cite{sanders_convivial_2013}. Themes were concluded by the two researchers independently and agreed in a post-coding meeting with a third researcher. Details of the themes can be found in the feedback section (Section 4.5.5). There was no limit for the length of participants’ responses.

\subsection{Apparatus and Setup}
We used an Oculus Rift CV1 as our VR HMD and a 50-inch 4K TV as our LD. Both devices were connected to an HP Z workstation with an i7 CPU, 16GB RAM, and a Nvidia Quadro P5200 GPU. Players’ gestures were detected via a Microsoft Kinect 2, which was also connected to the HP Z workstation. The heart rate (HR) was monitored by a Polar OH1 optical HR sensor, which has been proven to be reliable compared to the gold standard of HR measurement with an electrocardiography device \cite{hettiarachchi_validation_2019,schubert_polar_2018}. Figure 2 shows the experiment setup and devices used in the experiment.

\begin{figure}[h]
  \centering
  \includegraphics[width=0.6\linewidth]{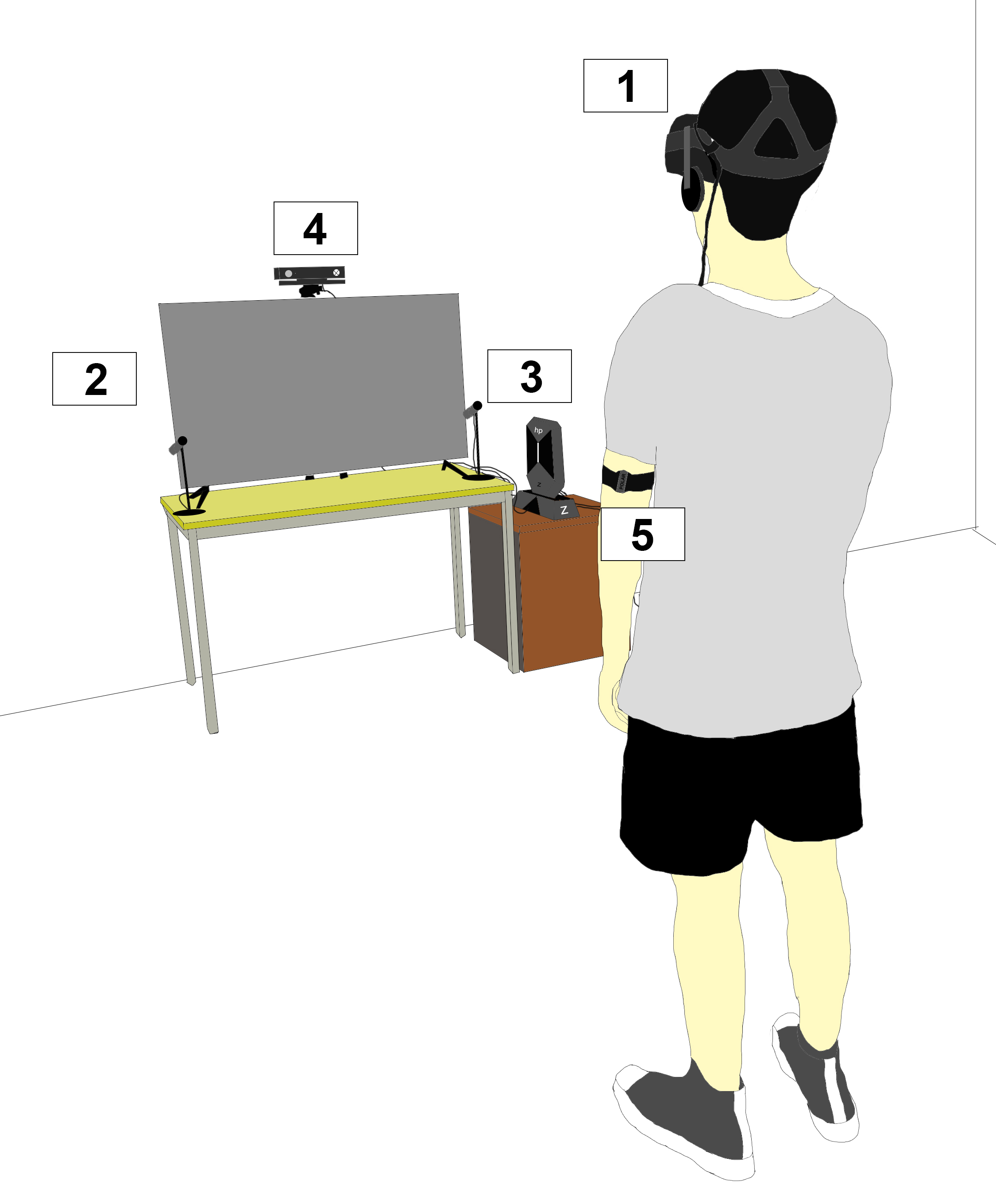}
  \caption{Experiment setup and the devices used in the experiment: (1) the Oculus Rift CV1; (2) a 50-inch 4K TV; (3) the HP Z backpack; (4) the Microsoft Kinect 2; and (5) Polar OH1.}
  \Description{Experiment setup}
\end{figure}

The experiment was conducted in an indoor laboratory room that could not be seen from the outside. The laboratory room was well illuminated, and its temperature was controlled by an air conditioner that regulated the room temperature to 24℃ during the experiment. 

\subsection{Participants}
\subsubsection{Inclusion and Exclusion Criteria}
Participants were recruited from a local university campus and a local community center through posters, social media platforms, and a mailing list for young adults between 18 and 30 years old and middle-aged adults between 45 to 65 years old. The study included participants who were not disabled, were not pregnant (because of the physical exertion required to play the game), and had not consumed any alcohol during the day (because blood alcohol level of approximately 0.07\% could reduce symptoms of cybersickness \cite{iskenderova_drunk_2017}, which might affect the results of our study). 

Participants were excluded from the experiment if they (1) answered “yes” to any of the Physical Activity Readiness Questionnaire \cite{thomas_revision_1993} questions, (2) had resting blood pressure higher than 140/90 mmHg, and (3) had an extremely good or poor resting heart rate (RestHR) level (i.e., heart rate range were the top 10\% or the last 10\% of the population) depending on their age and gender \cite{ostchega_resting_2011}. 

\subsubsection{Participants Background}
Thirty-two (32) participants participated in our study\textemdash 16 young adults (6 females; mean age = 20.6, SD = 1.31, range 18 to 23; BMI = 20.3, SD = 2.62), and 16 middle-aged adults (5 females; mean age = 47.7, SD = 2.68, range 45 to 54; BMI = 23.8, SD = 2.04). Among young adults, 7 of them had experience with VR HMDs, but none were regular users. Fourteen of them played videogames before; 6 of them played regularly. For middle-aged adults, none had experience with VR HMDs and videogames. There were no dropouts in this experiment. 

\subsection{Procedure and Task}
The duration of each session was about one hour. Before the experiment began, participants needed to fill out a pre-experiment questionnaire that gathered demographic information (e.g., age, gender, and experience with the VR device) and Physical Activity Readiness Questionnaire \cite{thomas_revision_1993}. After a brief description of the experimental procedure, participants signed the consent to participate in the experiment and collected their RestHR and resting blood pressure level. They were also asked to enter their age, gender, height, and weight into the Polar Beat app.

Before each condition started, a researcher would help each participant to wear the required devices (e.g., Polar OH1). Once their HR reached the equivalent RestHR level, they were led to the experiment stage, beginning with a training (warm-up) phase and then the gameplay phase (see Figure 1 and Section 3.2). After each condition, they were asked to fill in post-condition questionnaires (GEQ \cite{ijsselsteijn_game_2008}, SSQ \cite{kennedy_simulator_1993}, Borg RPE 6-20 scale \cite{borg_psychophysical_1982}). They proceeded to the next condition when they felt rested and their HR was at the resting level. Once they completed all conditions, they needed to complete a post-experiment questionnaire and a semi-structured interview.

\subsection{Results}
\subsubsection{Statistical Analysis}
We used SPSS version 24 for windows for data analysis. We employed a three-way mixed ANOVA with GC (uncertain and certain) and DT (VR and LD) as within-subjects variables and Age (young adults\textemdash YA and middle-aged adults\textemdash MA) as the between-subjects variable. We applied Age as the between-subjects variable because we want to follow existing approaches in the literature \cite{Nacke_BrainTraining,gerling_is_2013,wang_age-related_2017}. Bonferroni correction was used for pairwise comparisons. Effect sizes ($\eta_{p}^{2}$) were added whenever feasible. To minimize any impact on the readability of the paper, we have placed all the data results in the tables of an appendix located after the references. 

\subsubsection{Performance}
\textit{Completion Time on Each Life}. Figure 3a presents the mean completion time of each life (i.e., monster’s life1, life2, life3). ANOVA tests yielded a significant effect of Age on life2 ($F_{1,30} = 7.246, p < .05, \eta_{p}^{2} = .195$) and life3 ($F_{1,30} = 9.088, p < .01, \eta_{p}^{2} = .232$). Post-hoc pairwise comparisons revealed that YA could destroy the monster faster than MA on life2 and life3. No other significant effects were found.

\begin{figure*}
  \centering
  \includegraphics[width=\textwidth]{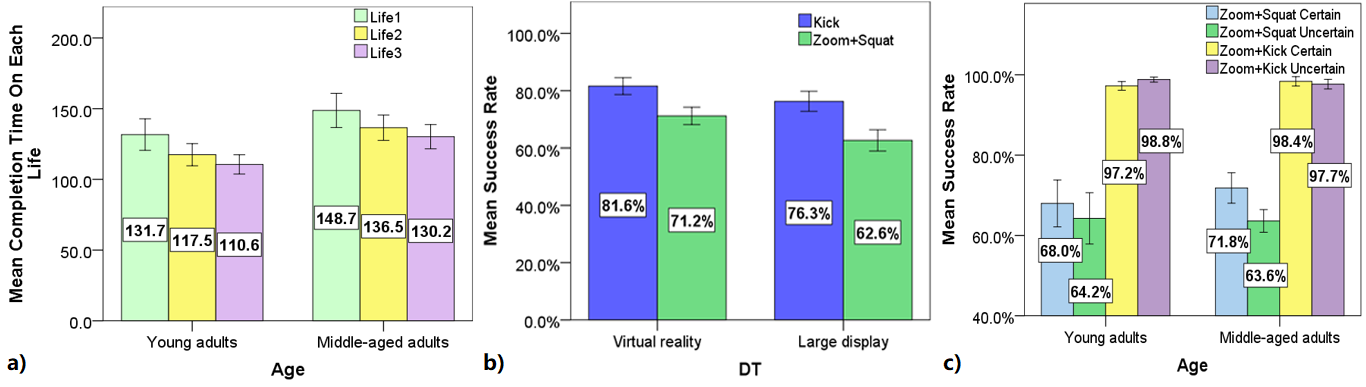}
  \caption{(a) Mean completion time on each monster’s life according to age group, (b) mean success rate of \textit{Kick} and \textit{Zoom}+\textit{Squat} according to DT, and (c) mean success rate of \textit{Zoom}+\textit{Squat} and \textit{Zoom}+\textit{Kick} according to GC and Age. Error bars indicate ±2 standard errors.}
  \Description{Fig3}
\end{figure*}

\textit{Success Rate}. Table 2 shows the ANOVA tests of the success rate for \textit{Zoom}+\textit{Squat}, \textit{Kick}, \textit{Zoom}+\textit{Kick}. Corresponding success rate data can be found in Figure 3b,c and Figure 4a. In summary, (1) participants have a higher defense (i.e., \textit{Zoom}+\textit{Squat}) success rate in certain GC than uncertain GC, (2) YA have a higher defense success rate in VR than LD, (3) participants have a higher \textit{Kick} success rate in VR than LD, (4) YA had a higher \textit{Zoom}+\textit{Kick} success rate than MA in VR, (5) YA had a higher \textit{Zoom}+\textit{Kick} success rate in VR than LD, and (6) YA had a higher \textit{Zoom}+\textit{Kick} success rate than MA in uncertain GC.

\textit{Total Number of Gestures Performed}. Table 3 shows the ANOVA tests of the total number of gestures performed for \textit{Zoom}+\textit{Squat}, \textit{Punch}, \textit{Zoom}+\textit{Kick}. Corresponding success rate data can be found in Figure 4b,c. In summary, (1) YA and MA both performed more defense moves (i.e., \textit{Zoom}+\textit{Squat}) in uncertain GC than certain GC, (2) MA performed more defense moves than YA in both certain and uncertain GC, (3) YA performed more \textit{Punch} than MA in LD, (4) MA performed more \textit{Punch} in VR than LD, (5) participants performed more \textit{Zoom}+\textit{Kick} in uncertain GC than in certain GC. 

\begin{table*}
  \caption{Three-way mixed ANOVA test results for success rate. Significant results where $p < .05$ are shown in light green, $p < .01$ in green, and $p < .001$ in dark green. \textit{Punch}, Age, DT × GC, DT × Age × GC have no significant results and therefore not shown for better clarity. No sig indicates no significant results.}
  \label{tab:Table2}
    \centering
    \begin{tabular}{p{1.2cm} p{4.7cm} p{5.2cm} p{5.4cm}}
  \hline
     & {\textit{Kick}} &   {\textit{Zoom}+\textit{Squat}}  & {\textit{Zoom}+\textit{Kick}}   \\
     
  \hline
  DT &\cellcolor{green!8} $F_{1,30} = 4.836, p < .05, \eta_{p}^{2} = .139$ & \cellcolor{green} $F_{1,30} = 14.403, p < .001, \eta_{p}^{2} = .324$  & No sig \\
  
  GC & No sig  & \cellcolor{green} $F_{1,30} = 21.799, p < .001, \eta_{p}^{2} = .421$ &  No sig \\
  
  DT × Age & No sig  &\cellcolor{green!30}  $F_{1,30} = 7.942, p < .01, \eta_{p}^{2} = .209$  &\cellcolor{green!8} $F_{1,30} = 5.008, p < .05, \eta_{p}^{2} = .143$  \\
  
  GC × Age & No sig   & No sig & \cellcolor{green!8} $F_{1,30} = 6.439, p < .05, \eta_{p}^{2} = .177$ \\
  
  Post-hoc & \textbf{DT}: VR > LD ($p < 0.5$; see Figure 3b)  & \textbf{GC}: uncertain < certain ($p < .001$; see Figure 3c);
\textbf{YA}: VR > LD ($p < .001$; see Figure 4a)
   &  \textbf{VR}: YA > MA ($p < .05$; see Figure 4a);
\textbf{YA}: VR > LD ($p < .05$; see Figure 4a);
\textbf{Uncertain}: YA > MA ($p < .05$; see Figure 3c)
 \\
  \hline
  \end{tabular}%
  \centering
\end{table*}

\begin{table*}
  \caption{Three-way mixed ANOVA test results for the total number of gestures performed. Significant results where $p < .05$ are shown in light green, $p < .01$ in green, and $p < .001$ in dark green. \textit{Kick}, DT, GC × DT, Age × GC × DT have no significant results and therefore not shown for better clarity. No sig indicates no significant results.}
  \label{tab:Table3}
    \centering
    \begin{tabular}{p{1.2cm} p{4.7cm} p{5.2cm} p{5.4cm}}
  \hline
     & {\textit{Punch}} &   {\textit{Zoom}+\textit{Squat}}  & {\textit{Zoom}+\textit{Kick}}   \\
     
  \hline
  GC & No sig  & \cellcolor{green} $F_{1,30} = 129.718, p < .001, \eta_{p}^{2} = .812$ & \cellcolor{green!8} $F_{1,30} = 5.473, p < .05, \eta_{p}^{2} = .154$ \\
  
  Age & \cellcolor{green!8} $F_{1,30} = 5.268, p < .05, \eta_{p}^{2} = .149$     & \cellcolor{green} $F_{1,30} = 18.638, p < .001, \eta_{p}^{2} = .383$ &  No sig \\
  
  GC × Age & No sig   &\cellcolor{green!30}  $F_{1,30} = 9.231, p < .01, \eta_{p}^{2} = .235$  & No sig \\
  
  DT × Age & \cellcolor{green!8} $F_{1,30} = 4.981, p < .05, \eta_{p}^{2} = .142$   & No sig &  No sig \\
  
  Post-hoc & \textbf{LD}: YA > MA ($p < .01$; see Figure 4b);
\textbf{MA}: VR > LD ($p < .01$; see Figure 4b)
  &  \textbf{YA and MA}: uncertain > certain (both $p < .001$; see Figure 4c);
\textbf{Uncertain and certain}: MA > YA (both $p < .001$; see Figure 4c)
   &  \textbf{GC}: uncertain > certain ($p < .05$; see Figure 4c) \\
  
  \hline
  \end{tabular}%
  \centering
\end{table*}

\begin{figure*}
  \centering
  \includegraphics[width=\textwidth]{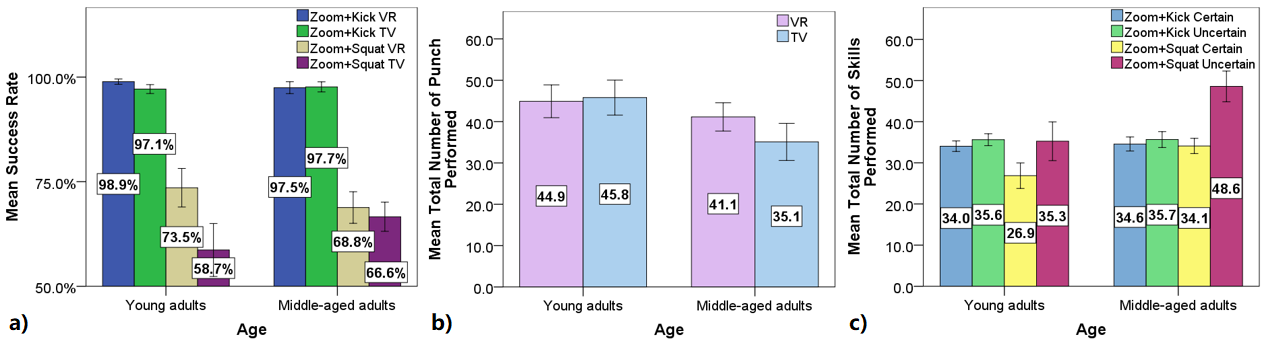}
  \caption{(a) Mean success rate of \textit{Zoom}+\textit{Kick} and \textit{Zoom}+\textit{Squat} according to DT and Age, (b) mean total number of \textit{Punch} performed according to DT and Age, and (c) mean total number of \textit{Zoom}+\textit{Kick} and \textit{Zoom}+\textit{Squat} performed according to GC and Age. Error bars indicate ±2 standard errors.}
  \Description{Fig4}
\end{figure*}

\subsubsection{Experience}
\textit{Game Experience}. ANOVA tests yielded a significant effect of Age on competence ($F_{1,30} = 20.787, p < .001, \eta_{p}^{2} = .409$), immersion ($F_{1,30} = 23.010, p < .001, \eta_{p}^{2} = .434$), tension ($F_{1,30} = 20.815, p < .001, \eta_{p}^{2} = .410$), negative affect ($F_{1,30} = 19.278, p < .001, \eta_{p}^{2} = .391$), positive affect ($F_{1,30} = 20.810, p < .001, \eta_{p}^{2} = .410$). Post-hoc pairwise comparisons showed that YA had a higher levels of competence, immersion, tension, negative affect, and positive affect than MA (see Figure 5a).

\begin{figure*}
  \centering
  \includegraphics[width=\textwidth]{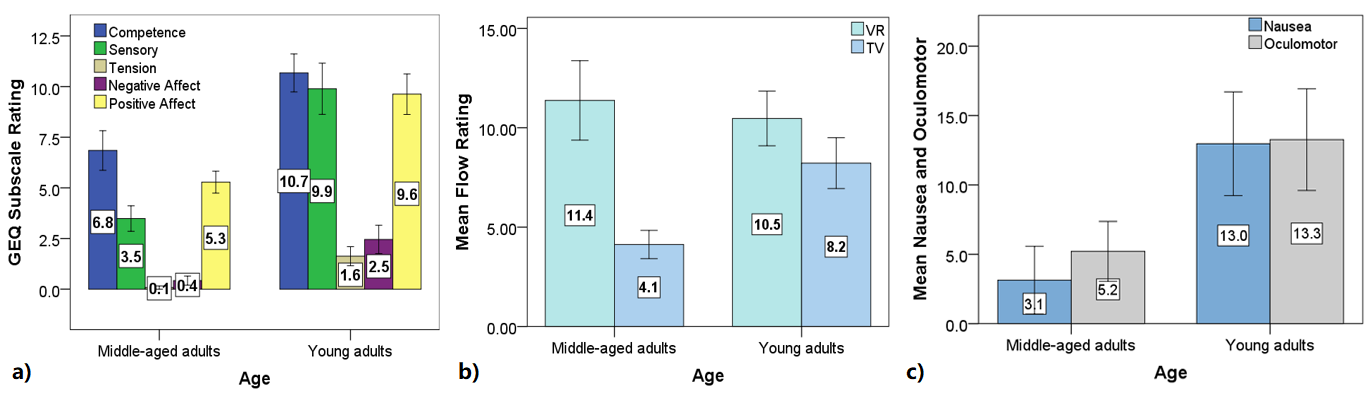}
  \caption{(a) Game experience questionnaire rating of subscales according to Age, (b) mean flow rating according to DT and Age, and (c) mean nausea and oculomotor rating according to Age. Error bars indicate ±2 standard errors.}
  \Description{Fig5}
\end{figure*}

There was a significant effect of DT ($F_{1,30} = 40.298, p < .001, \eta_{p}^{2} = .573$) on flow, showing that participants experienced a greater flow in VR than LD. Additionally, ANOVA tests yielded a significant effect of DT × Age ($F_{1,30} = 11.163, p < .01, \eta_{p}^{2} = .271$) on flow. Post-hoc pairwise comparisons revealed that (1) YA experienced a lower flow than MA in LD ($p < .001$), (2) VR could lead to a greater flow experience than LD in both YA ($p < .05$) and MA ($p < .001$). Figure 5b depicts the corresponding flow values. No other significant effects were found.

\textit{Simulator Sickness}. ANOVA tests yielded a significant effect of Age on nausea ($F_{1,30} = 7.049, p < .05, \eta_{p}^{2} = .190$) and oculomotor ($F_{1,30} = 5.242, p < .05, \eta_{p}^{2} = .149$), but not on disorientation ($F_{1,30} = 2.490, p = .125, \eta_{p}^{2} = .077$). Post-hoc pairwise comparisons revealed that (1) YA experienced a higher nausea level than MA (see Figure 5c), and (2) YA experienced a higher oculomotor level than MA (see Figure 5c). No other significant effects were found.

\textit{Uncertain Elements’ Ratings}. We employed a two-way mixed ANOVA with Elements (false-attack, hit, miss) as the within-subjects variable and Age as the between-subjects variable. The ANOVA tests yielded a significant effect of Elements ($F_{1.607,48.224} = 3.547, p < .05, \eta_{p}^{2} = .106$), but not Elements × Age ($F_{1.607,48.224} = 1.656, p = .200$) on the ratings of the uncertain elements. There was a significant effect of Age ($F_{1,30} = 8.217, p < .001, \eta_{p}^{2} = .215$) on the uncertain elements’ ratings, showing that uncertainty settings were rated higher in YA (M = 5.88, s.e. = 0.20) than MA (M = 5.08, s.e. = 0.20). However, post-hoc pairwise comparisons could not find any significance between uncertain elements.

\subsubsection{Exertion}
Table 4 shows the ANOVA tests of all exertion measures. In summary, (1) YA had lower avgHR\% than MA in uncertain GC, (2) MA had a higher avgHR\% in uncertain GC than certain GC, (3) participants burned more calories in uncertain GC than certain GC, (4) MA participants burned more calories than YA participants (see Figure 6b), (5) Borg RPE for uncertain GC was higher than certain GC among YA and MA, (6) the Borg RPE for YA was higher than MA in certain GC and uncertain GC. 

\begin{table*}
  \caption{Three-way mixed ANOVA test results for exertion measurements. Significant results where $p < .05$ are shown in light green, $p < .01$ in green, and $p < .001$ in dark green. DT, GC × DT, Age × DT, Age × GC × DT have no significant results and therefore not shown for better clarity. No sig indicates no significant results.}
  \label{tab:Table4}
    \centering
    \begin{tabular}{p{1.2cm} p{4.7cm} p{5.2cm} p{5.4cm}}
  \hline
     & {avgHR\%} &   {Calories Burned}  & {Borg RPE}   \\
     
  \hline
  GC & \cellcolor{green} $F_{1,30} = 30.560, p < .001, \eta_{p}^{2} = .505$  & \cellcolor{green} $F_{1,30} = 45.587, p < .001, \eta_{p}^{2} = .603$ & \cellcolor{green} $F_{1,30} = 39.533, p < .001, \eta_{p}^{2} = .569$  \\
  
  Age & \cellcolor{green!30}$F_{1,30} = 7.754, p < .01, \eta_{p}^{2} = .205$   & \cellcolor{green!30}$F_{1,30} = 8.353, p < .01, \eta_{p}^{2} = .218$ & \cellcolor{green} $F_{1,30} = 15.488, p < .001, \eta_{p}^{2} = .340$ \\
  
  GC × Age & \cellcolor{green!30}$F_{1,30} = 8.279, p < .01, \eta_{p}^{2} = .248$  & No sig &\cellcolor{green!8} $F_{1,30} = 4.759, p < .05, \eta_{p}^{2} = .137$ \\
  
  Post-hoc & \textbf{Uncertain}: YA < MA (both $p < .01$; see Figure 6a);
\textbf{MA}: uncertain > certain ($p < .001$; see Figure 6a)
 & \textbf{GC}: uncertain > certain ($p < .001$; see Figure 6b);
\textbf{Age}: MA > YA ($p < .01$; see Figure 6b)
 & \textbf{YA}: uncertain > certain ($p < .01$; see Figure 6c);
\textbf{MA}: uncertain > certain ($p < .001$; see Figure 6c);
\textbf{Certain}: YA > MA ($p < .001$; see Figure 6c);
\textbf{Uncertain}: YA > MA ($p < .01$; see Figure 6c)
 \\
  
  \hline
  \end{tabular}%
  \centering
\end{table*}

\begin{figure*}
  \centering
  \includegraphics[width=\linewidth]{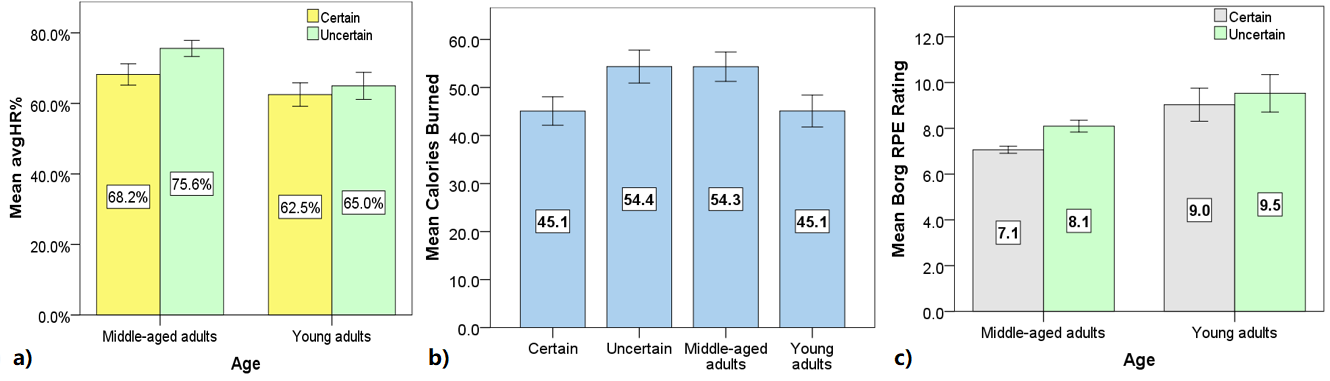}
  \caption{(a) Mean avgHR\% according to GC and Age, (b) mean calories burned, and (c) mean Borg RPE rating according to GC and Age. Error bars indicate ±2 standard errors.}
  \Description{Fig6}
\end{figure*}

\subsubsection{User Rankings and Feedback}
The VR uncertain version was rated the best version among the four versions by 23 participants (12 YA). Only 5 participants (4 YA) selected VR certain as their top option and 4 MA chose LD uncertain version as their top selection. 

\textit{Feedback}. From the coded transcripts, three main themes emerged (element of the games, general gaming experience, and exercising for health) from the two researchers, who first reviewed the transcripts independently. They were agreed by a third researcher after a second discussion. Thirty-two participants were labeled P1-P16 (YA group) and P17-P32 (MA group). 

Overall, both user groups perceived the game as “\textit{enjoyable}” (10 YA, 9 MA), “\textit{novel}” (9 YA, 8 MA), and “\textit{good for their health}” (9 YA, 14 MA) and none of them perceive anything that was confusing in the game. Both groups perceived the false-attacks more difficult than expected (P3, P13, P20, P22, P24-27), but only MA participants mentioned that sometimes they could not perform the defense move in time.

Regarding the elements that they liked about the game, the comments from the two groups came from two different perspectives. Most YA focused on the game elements (e.g., “\textit{the false-attack by the opponents}” [P3, P14, P16], “\textit{critical hits}” [P5], “\textit{misses}” [P11], “\textit{using gestures to trigger attacks are fun and easy to understand}” [P6, P9, P13]) while only a few mentioned about the health benefits as their preferred elements (P8, P10, P15). This is a completely different for the MA, where 13 MA mentioned they liked the game because it could be a good exercise activity while only 6 comments focused on design elements (e.g., “\textit{false-attacks by the monster is a good design}” [P23, P27, P30], “\textit{it tricks me into performing defense moves, which is good for my health}” [P20, P24, P25]).

The two generations focused on the different perspectives again regarding the elements that they did not like. Most comments from YA were about the graphics and models used in the game, that they should be improved and more moves could be added. On the other hand, most MA believed that the uncertain elements are sometimes overused, which caused them to perform too many defense moves and made them feeling exhausted during the game.

\section{Discussion}
\subsection{Effect of Age on Exergames}
In general, the performance (i.e., completion times, success rates for both attack and defense moves) of middle-aged adults were worse than young adults in our motion-based first-person exergame, which is in line with previous studies of similar games \cite{gerling_is_2013}. One possible reason could be age-related declines in mobility; for instance, middle-aged adults typically require more time to perform gestures \cite{ferrucci_age-related_2016}. They also were not able to react to the monster’s attack sometimes or cancel their defense moves when realizing that the monster was performing false-attacks; for example, P20, P22, P24-25, P27-28: “\textit{I could not react in time}.” Hence, it is necessary to take into account age-related declines (e.g., working memory \cite{meguro_nature_2000}, grip strength \cite{kozakai_sex-differences_2016}, and muscle mass \cite{brown_complexity_1996}) when designing exergames for middle-aged adults.

In addition, the two age groups perceived the game experience differently. We found that young adults were more immersive (immersion, flow) in the game than middle-aged adults and had a higher positive emotion, efficacy, competence. However, young adults still felt more annoyed and experienced more negative emotions than middle-aged adults even though they had a better performance (e.g., the successful attack rate is much higher). One possible reason is that young adults might have expected that they should perform much better due to their competitive expectations of themselves and the game, while the competition was downplayed in middle-aged adults \cite{subramanian_assessing_2019}.

Previous research has suggested that there may be a decline in susceptibility to VR sickness as people age \cite{bardey_motion_2013}. Our results also support this, as we found that young adults felt sicker during gameplay than middle-aged adults. Overall, sickness level for all participants were either negligible or very low, with no participants experiencing severe simulator sickness. That is, all participants had no issues in playing the game.

Existing literature in the exercise domain (e.g., tai chi \cite{lan_relative_2004}, arms training \cite{groslambert_effects_2006}, arm abduction \cite{pincivero_rpe_2010}) have suggested that age does not affect the exertion level of the exercise. However, this is not supported by our results because we found that our two groups of participants produced different levels of exertion (middle-aged adults had a higher avgHR\% in the uncertain condition and burned more calories than young adults but gave a lower Borg RPE ratings). Further study is required to explain this.

\subsection{Effect of Display Type on Exergames}
Our results suggest that participants had a better performance in VR (i.e., higher success rates in attack and defense moves in VR than LD). This is understandable because the greater flow experience brought by VR to the players had a positive effect on performance in the game \cite{admiraal_concept_2011}. A previous study \cite{xu_studying_2020} that also focused on the effect of DT versus VR showed that VR could provide a greater positive game experience (e.g., challenge, flow, immersion) to the players than LD, which was also found in our results (i.e., VR led to a higher flow rating than LD). Existing literature also indicated that game experience (from GEQ) could be perceived the same in both VR and LD \cite{xu_assessing_2019}. One reason could be that in \cite{xu_assessing_2019}, participants only experienced 4 minutes of gameplay, which is relatively short for developing a fuller picture of the technologies. Hence, we suggest that future studies consider a longer game duration, like 7- 8 minutes used in our research and in \cite{xu_studying_2020}, to let the players experience a game in each technology more fully.

In addition, our findings indicate there was no significant difference regarding the level of sickness that participants experienced between VR and LD when playing the motion-based exergame, which is in line with \cite{xu_assessing_2019} but not \cite{xu_studying_2020} where researchers reported that playing a motion-based exergame in VR could lead to a higher sickness than LD. One possible explanation could be that the type of game used in the experiment was different. Our game and the game used in \cite{xu_assessing_2019} involved more interaction with the virtual world than the game in \cite{xu_studying_2020}. For instance, players had direct contacts with the virtual objects (either through attacking and defending against the monster in our game or directly using the hands or feet to hit the objects in the game from \cite{xu_assessing_2019}), which is not the case for \cite{xu_studying_2020} where the gestures performed by the users did not have direct contact with the virtual objects in the form of cubes.

\subsection{Effect of Uncertainty on Exergames}
The purpose of the design of false-attacks, one uncertain element in our exergame, was to trick the players into using the defense moves. Our results show that this element achieved its intended goal because participants performed more defense moves (\textit{Zoom}+\textit{Squat}) in the uncertain condition than the counterpart condition. We also observed during the experiment that this design tricked all players across both groups.

In addition, the design of misses had also forced them to perform more attack moves in their attempts to kill the monster. Hence, participants had a higher exertion level (i.e., avgHR\%\textemdash MA, calories burned, Borg RPE) in the uncertain condition. Furthermore, what is interesting to note is that participants did not feel a worse experience by these design features since (1) they did not complain about the features, and (2) the gameplay experience and sickness in both game conditions were not significantly different. Therefore, we believe that involving uncertain elements (i.e., false-attacks, misses, and critical hits) in the type of exergame similar to ours could increase players' energy costs without incurring negative gameplay experiences in both VR and LD.

\subsection{Design Guidelines}
\subsubsection{Applying Uncertainty to Exergames}
As our results show, the proposed uncertain elements in our exergame could be useful in enhancing exertion levels during game sessions. We list with examples of how these uncertain elements can be applied to other exergames.

For sports exergames, false-attack can be used in several ways. For example, in the boxing game \textit{Creed: Rise to Glory}\footnote{https://www.oculus.com/experiences/rift/1872428116153565/}, a false-attack can be directly applied to Creed's attack strategy to trick players into making defense moves. False-attacks can be enhanced further by following a real attack after the animation of a false-attack. For \textit{Eleven Table Tennis VR}\footnote{https://www.oculus.com/experiences/rift/989106554552337/}, this can be added as a way for NPC to pretend they want to move into one direction but not moving into that direction. This type of false moves can be used in designing basketball and football exergames where trickery is a key to make a defending player go into one direction so that the player can move into the opposite way (e.g., \textit{Kinect Sports: Soccer}\footnote{https://marketplace.xbox.com/en-US/Product/Kinect-Sports/66acd000-77fe-1000-9115-d8024d5308c9}).

For exergames that involve one-way interaction with the enemy (i.e., player to NPC), critical hits and misses can be used. For instance, in the tower defense game \textit{Longbowman} \cite{Exercise_intensity_driven}, critical hits and misses can be designed with additional features. A critical hit can deal additional damage and also slow down the movement of the enemy. In contrast, a miss does not damage the enemy and would make the enemy become angry and move faster.

For exergames that involve two-way interaction with the enemy (player to NPC and NPC to player), all three elements can be used. For instance, in \textit{Ring Fit Adventure}, a motion-based active game for the Nintendo Switch, all these elements can be added in a similar way that we did in our exergame since it is designed based on this commercial game.

\subsubsection{Highlighting Health Benefits to Middle-aged and Older Adults}
Like older adults \cite{subramanian_assessing_2019}, middle-aged adults believe that exergames are helpful to their health. We suggest making the potential health benefits to middle-aged adults explicit and clear inside the game and as part of the gameplay experience. For instance, designers could (1) introduce the benefits of each gesture before the game, (2) present the energy cost like calories burned during the game as part of any dynamic visual and audio feedback, (3) give a summary report of the overall performance (e.g., for each type of gestures, providing the total number the player performed) after the game.

\subsection{Limitations and Future Work}
There are some limitations in this research, which can also serve as directions for the future. One limitation is that we tested three elements of uncertainty (false-attacks, misses, and critical hits) that covers four uncertainty sources. Future work could explore more uncertainty sources \cite{costikyan_uncertainty_2013} in motion-based exergames. For example, we can use (1) \textit{analytical complexity}, by allowing more skills for the player but require the player to kill the monster in a limited time so that the player needs to analyze the best strategy to fight against their opponent carefully. It is possible to integrate (2) \textit{hidden information}, by not showing information of the opponent’s attack moves. Addition, (3) \textit{narrative anticipation} can be used by adding a storyline to a game and fighting an opponent would reward them with the corresponding piece of the storyline. By doing this, the player has the desire to know the next piece of the storyline \cite{murnane_designing_2020}. 

In addition, there are some limitations related to the choice of VR HMD and exergames in current commercial VR HMDs. We used the Oculus Rift CV1. Newer VR HMDs (i.e., VIVE Pro Eye) that come with a higher resolution could impact simulator sickness and game experience. We used the Oculus Rift CV1 because we wanted to have consistency with prior studies \cite{xu_assessing_2019,xu_studying_2020}. The Rift CV1, as a tethered helmet, has a limited range of motion because of the attached cables. While standalone devices like Oculus Quest do not have this limitation, they suffer from latency issues when used with external motion sensors (i.e., Kinect) to capture motion data. In addition, long gameplay sessions wearing any current HMDs could result in sweats in the glasses; thus, the length of gameplay should be carefully designed to prevent this issue. Also, to make MA-friendly exergames, future games should involve more simple gestures (like zoom\textemdash hands stretching out, hands-up) to eliminate any risks when wearing a VR HMD.

Our study only involved a single session. Longer-term studies will be useful to determine if the same results hold and to determine additional effects that may come with long-term exposures. In addition, due to the COVID-19, we cannot to include the elderly adults (i.e., those 65 years old and above) in the experiment. Future work could have all these three groups of adults (i.e., young, middle-aged, elderly) to assess their relative performance and experience with exergames. 

\section{Conclusion}
In this research, we have investigated the effect of display type (virtual reality and large display) with or without elements of uncertainty in motion-based first-person perspective exergames. We also have explored the impact of age by comparing game performance, gameplay experience, and level of energy exertion between young adults and middle-aged adults. Our results suggest the following three conclusions: (1) For the type of exergame like ours, virtual reality could improve game performance while maintaining the same level of sickness as large displays. (2) Uncertain elements like those used in this research's motion-based exergame might not help enhance the overall game experience, but are instrumental in increasing exertion levels, which is one of the essential features of exergames. (3) Exergames for middle-aged adults should be carefully designed with consideration to age-related declines, similar to older adults. We also proposed two main design guidelines which can pave the way for improving the acceptability of VR exergames among young and middle-aged adults.

\begin{acks}
The authors would like to thank the anonymous reviewers for their valuable comments and helpful suggestions and the Committee Member who guided the revision of our paper. The work is supported in part by Xi’an Jiaotong-Liverpool University (XJTLU) Key Special Fund (KSF-A-03) and XJTLU Research Development Fund.
\end{acks}

\bibliographystyle{ACM-Reference-Format}
\balance 


\appendix

\section{Appendix}
\subsection{Data Results}
We list all data in Table 5 and 6. VR\_Cer represents VR certain game condition (GC), VR\_Unc represents VR uncertain GC, TV\_Cer represents TV certain GC, TV\_Unc represents TV uncertain GC. 

\begin{table*}
  \caption{Means (SDs) of participants' performance data regarding the completion time on each of the three lives of the monster, total number of gestures performed, and success rate of each move.}
  \label{tab:Table5}
    \centering
    \begin{tabular}{p{1.8cm} p{1.6cm} p{1.6cm}  p{1.6cm} p{1.6cm}  p{1.6cm} p{1.6cm}  p{1.6cm} p{1.6cm}}
  \hline
  & \multicolumn{4}{c} {Young Adults} & \multicolumn{4}{c} {Middle-aged Adults}   \\
  
  \hline
    Type & {VR\_Cer} & {VR\_Unc}  & {TV\_Cer} & {TV\_Unc}  & {VR\_Cer}  & {VR\_Unc} & {TV\_Cer} & {TV\_Unc}  \\
  \hline
 \multicolumn{9}{c} {Completion Time on Each of The Three Lives of The Monster}\\
 \hline
    Life1 & 126.83 (34.74)	& 126.14 (36.79) &	133.69 (54.42)	& 140.05 (51.73) &	149.38 (41.71)	& 152.82 (48.91)	& 148.16 (60.72)	& 144.64 (44.30)  \\
  
 Life2 & 113.80 (22.07)	& 114.60 (33.05)	& 119.71 (33.22) &	121.71 (37.82) &	133.96 (25.06) &	138.09 (39.38) &	131.36 (29.37) &	142.75 (47.57) \\
  
 Life3 & 105.45 (19.23)	& 109.30 (35.73) &	112.21 (29.10) &	115.39 (23.88) &	134.27 (36.88)&	126.20 (28.26) &	121.08 (19.63) &	139.17 (46.41) \\
 \hline
 \multicolumn{9}{c} {Total Number of Gestures Performed}\\
 \hline
     Kick &	33.19 (7.88) &	35.13 (7.44) &	38.00 (8.33) &	34.75 (10.08)	& 36.50 (10.41)	& 36.56 (8.60)	 & 35.19 (9.09)	 & 35.00 (9.35) \\
  
 Push	&45.56 (13.77)&	44.25 (8.31)	&47.50 (9.64)&	44.13 (14.06)	&40.94 (10.97)	&41.31 (8.54)	&34.31 (13.80)	&35.88 (11.91) \\
  
 Zoom+Kick&	35.25 (3.62)&	35.75 (5.01)&	32.81 (3.29)	&35.50 (3.12)&	34.06 (4.55)	&35.75 (5.42)&	35.06 (5.20)&	35.56 (5.67) \\
 
 Zoom+Squat	&29.19 (7.88)	&33.88 (13.50)	&24.56 (9.24)	&36.63 (13.44)	&33.75 (4.93)	&46.69 (7.43)	&34.44 (5.68)	&50.44 (12.93)\\
 \hline
  \multicolumn{9}{c} {Success Rate of Each Move}\\
 \hline
 Kick	&82.19\% (12.31\%)&	83.72\% (10.28\%)&	77.28\% (13.50\%)&	75.87\% (21.27\%)&	80.03\% (12.01\%)&	80.44\% (13.10\%)&	75.96\% (11.01\%)	&76.04\% (7.98\%)\\

Push &	52.04\% (25.06\%)	&55.34\% (24.32\%)	&61.05\% (18.84\%)&	60.43\% (20.57\%)&	54.76\% (13.12\%)	&49.18\% (18.52\%)&	56.79\% (15.61\%)&	55.31\% (15.83\%)\\

Zoom+Kick	&98.32\% (2.14\%)&	99.50\% (1.08\%)&	96.16\% (3.46\%)&	98.13\% (2.17\%)&	97.61\% (4.08\%)&	97.31\% (4.03\%)&	99.14\% (2.20\%)	&96.19\% (3.81\%)\\

Zoom+Squat	&74.07\% (13.17\%)	&73.03\% (13.36\%)	&61.88\% (17.56\%)&	55.46\% (18.25\%)	&73.51\% (10.20\%)	&64.12\% (9.21\%)&	70.13\% (11.33\%)&	63.10\% (6.78\%)\\
 \hline
  \end{tabular}%
  \centering
\end{table*}

\begin{table*}
  \caption{Means (SDs) of participants' experience and exertion data regarding each game experience questionnaire subscale, simulator sickness questionnaire subscale, and exertion measurement.}
  \label{tab:Table6}
    \centering
    \begin{tabular}{p{2.1cm} p{1.55cm} p{1.55cm}p{1.55cm}p{1.55cm}p{1.55cm}p{1.55cm}p{1.55cm}p{1.55cm}}
  \hline
     & \multicolumn{4}{c} {Young Adults} & \multicolumn{4}{c} {Middle-aged Adults}   \\
  \hline
    Type & {VR\_Cer} & {VR\_Unc}  & {TV\_Cer} & {TV\_Unc}  & {VR\_Cer}  & {VR\_Unc} & {TV\_Cer} & {TV\_Unc}  \\
  \hline
  \multicolumn{9}{c} {Game Experience Questionnaire}\\
 \hline
Competence&	12.00 (3.72)&	9.00 (4.52)&	10.25 (2.70)&	11.44 (3.44)&	6.69 (4.09)&	6.75 (4.60)&	7.38 (4.18)&	6.56 (2.92)\\

Immersion	&9.88 (5.64)&	10.44 (4.83)&	9.19 (4.86)&	10.06 (5.28)&	3.44 (2.45)&	3.50 (2.76)&	3.50 (2.90)&	3.50 (2.13)\\

Flow	&10.63 (4.11)&	10.31 (3.79)&	7.69 (3.52)	&8.75 (3.75)&	10.94 (5.63)	&11.81 (5.83)&	3.94 (1.98)&	4.31 (2.09)\\

Tension	&1.31 (1.45)&	1.88 (2.45)&	1.63 (1.59)&	1.69 (2.02)&	0.19 (0.54)&	0.06 (0.25)&	0.00 (0.00)&	0.06 (0.25)\\

Challenge&	6.13 (3.88)&	7.25 (2.59)	&6.69 (2.96)&	6.25 (3.04)	&6.06 (2.69)&	6.00 (2.76)&	5.44 (2.68)	&6.44 (2.58)\\

Negative Affect&	2.00 (2.19)&	2.56 (2.56)&	2.44 (2.76)	&2.81 (3.69)&	0.56 (1.21)&	0.63 (1.15)&	0.31 (0.60)&	0.19 (0.40)\\

Positive Affect&	10.56 (4.57)&	9.38 (3.90)&	8.88 (4.08)&	9.69 (3.59)&	4.94 (2.21)	&5.19 (2.14)&	5.38 (2.39)&	5.63 (2.09)\\
 \hline
 
 \multicolumn{9}{c} {Simulator Sickness Questionnaire}\\
 \hline
 Nausea	&11.93 (13.71)&	14.31 (17.42)&	11.93 (13.26)&	13.71 (16.33)&	2.98 (9.68)&	2.98 (8.33)&	1.19 (3.26)&	5.37 (14.77)\\

Oculomotor	&12.79 (14.57)&	11.84 (16.60)&	14.21 (13.80)&	14.21 (14.87)&	3.32 (7.81)	&6.16 (5.69)	&4.26 (9.17)&	7.11 (11.23)\\

Disorientation&	11.31 (23.41)&	6.96 (15.25)&	6.96 (21.56)	&7.83 (24.88)&	0.00 (0.00)&	0.00 (0.00)&	3.48 (9.51)&	1.74 (4.75)\\

Total SSQ &	14.03 (16.64)&	13.32 (17.43)	&13.56 (15.62)	&14.49 (18.16)	&2.81 (7.54)	&4.21 (5.27)	&3.51 (8.25)&	6.08 (12.13)\\
 \hline
 \multicolumn{9}{c} {Exertion}\\
 \hline
 avgHR\%&	62.61\% (8.87\%)&	66.12\% (11.72\%)	&62.39\% (10.15\%)&	63.79\% (10.14\%)&	67.75\% (7.22\%)&	76.26\% (6.12\%)	&67.83\% (8.73\%)&	74.92\% (6.90\%)\\

maxHR\%	&70.79\% (9.76\%)	&72.63\% (12.60\%)&	71.33\% (12.60\%)&	72.83\% (10.65\%)&	76.26\% (6.60\%)&	84.78\% (6.84\%)	&75.56\% (7.51\%)&	83.64\% (5.93\%)\\

Calories &	41.31 (10.73)&	50.88 (14.68)	&42.19 (14.79)&	46.06 (11.79)	&50.31 (10.26)&	60.25 (14.09)	&46.56 (9.61)&	60.25 (8.89)\\

Borg RPE 6-20	&8.88 (2.28)&	10.19 (2.71)	&9.19 (1.83)	&8.88 (1.67)	&7.06 (0.44)&	8.19 (0.75)	&7.06 (0.44)&	8.00 (0.73)\\
 \hline
  \end{tabular}%
  \centering
\end{table*}

\end{document}